\documentclass[12pt,a4paper]{article}
\usepackage{isolatin1}
\def\lsim{\mathrel{\vcenter{\hbox{$<$}\nointerlineskip\hbox{$\sim$}}}}
\def\gsim{\mathrel{\vcenter{\hbox{$>$}\nointerlineskip\hbox{$\sim$}}}}
\newcommand{\nc}{\newcommand}
\nc{\figcap}[1]{\begin{quote}\refstepcounter{figure}
        {\bf Figure \thefigure}: {\small #1}\end{quote}}

\begin{document}
\baselineskip=20 pt

\title{Neutrinos Confronting Large Extra Dimensions}
\author{J. Maalampi$^{1,2}$\thanks{jukka.maalampi@helsinki.fi}, \
 V. Sipil\"{a}inen$^3$\thanks{ville.sipilainen@helsinki.fi} \ and
 I. Vilja$^4$\thanks{vilja@utu.fi} \\
  $^1$\emph{Department of Physics, University of Jyväskylä,
                                              Finland} \\
$^2$\emph{Helsinki Institute of Physics, FIN-00014 University of Helsinki,
                                              Finland} \\
$^3$\emph{Department of Physics, FIN-00014 University of Helsinki,
                                              Finland} \\
$^4$\emph{Department of Physics, FIN-20014 University of Turku,
                                              Finland}}
\date{March 27, 2001}
\maketitle

\begin{abstract}
We study neutrino physics in a model with one large extra dimension.
We assume the existence of two four-dimensional branes in the
five-dimensional space-time, one for the ordinary particles and the other
one for mirror particles, and we investigate neutrino masses and mixings
in this scheme.
Comparison of experimental neutrino data with the
predictions of the model leads to various restrictions
on the parameters of the model. For instance,
the size of the extra dimension, $R$, turns out to be
bounded from below. Cosmological considerations
seem to favor a large $R$.
The usual mixing schemes proposed as solutions
to the solar and atmospheric neutrino
anomalies are compatible with our model.
\end{abstract}
\thispagestyle{empty}
\newpage
\section{Introduction}

The possibility that the physical space has more than three
dimensions \cite{0}
has attracted a great deal of attention recently.
Refs.\ \cite{1}--\cite{3} pioneered the idea that the
presence of extra compact spatial dimensions, some of which may
be as large as  fractions of millimeter, could lower the "fundamental"
Planck scale $M_\ast \, $, \emph{i.e.} the Planck scale in
the higher-dimensional space-time, down to the TeV energy range,
alleviating thereby the hierarchy problem. If the number of  new dimensions is
$n$ and we assume that all of them have the same radii of the size $R$, the
four-dimensional Planck scale $M_{Pl}$ is related to the fundamental Planck
scale through $M_{Pl}^2 \sim M_{\ast}^{n+2} R^n$.
Given the experimental fact that no deviations from the Newton law
has shown up down to  $R\simeq 1$ mm \cite{4}, it is immediately clear from this relation
that one extra dimension would not be enough for the solution of the
hierarchy problem. On the other hand, for $n=2$ one gets
$R \sim 0.1$ mm, which is in an interesting range
in view of the gravitational force experiments
discussed in \cite{4,5}.

The basic assumption of the extra dimension scenario
is that all the fields charged under the Standard
Model (SM) gauge group are localized on a brane, the familiar
3+1-dimensional space-time,
embedded in the 4+$n$-dimensional space, called the bulk \cite{1}.
Gravitons and other particles with no SM interactions are not confined to the brane
but are free to propagate in the bulk as well.
This brane-bulk structure is suggested by some string theoretical observations.

The large extra dimension scenario is particularly intriguing from the neutrino physics
point of view.
The right-handed chiral components of neutral leptons ($\nu_R$), if they
exist, are the only SM particles that can live in the bulk \cite{6,7}. The
Yukawa coupling between  $\nu_L$,  residing on the brane, and  $\nu_R$
is suppressed by a factor $M_\ast /
M_{Pl}$, providing a new and elegant explanation
for the lightness of neutrinos. Perhaps the most natural conventional explanation is due
to the seesaw mechanism, which requires
the existence of a new mass scale around $10^{12}$ GeV or higher.
In the extra dimension scheme such a high mass-scale is neither needed nor naturally
appears, but small neutrino masses follow
from the suppression of the Yukawa couplings due to the large bulk volume.

The existence of
the bulk neutrinos has another interesting consequence. The
bulk neutrinos  $\nu_{lR}\ (l=e,\mu ,\tau )$ appear in four dimensions as  Kaluza-Klein
excitations having Dirac mass terms with
the left-handed bulk components $\nu_{lL}$ that originate in the quantized
internal momenta in compact extra dimensions. The existence of these new
degrees of freedom enrich the neutrino
spectrum and the neutrino oscillation
patterns. Theoretical and phenomenological aspects  of this
scenario have been addressed in several recent papers
(see \emph{e.g.}\ \cite{9}--\cite{24b}).

The localization of the SM fields on the brane
may be explained in terms of non-perturbative effects in string theory
\cite{2}. An alternative and perhaps more intuitive approach is achieved in the
context of an effective field theory.
In this method one introduces a five-dimensional scalar
field (for $n=1$) with an effective potential that has a domain wall type profile
in the extra dimension \cite{25}--\cite{27}. The origin of the scalar field is usually left unspecified.
For example the so-called radion may be a natural,
though not the only, possibility for such a scalar field.
The radion field is associated to
the extra dimension components of the metric tensor
and therefore is always present in  the model.

The discussion in this paper is based on the
observation that the scalar
field responsible for the localization of the SM fermions
must form two kinks, \emph{i.e.}\ a kink and
an anti-kink (or any number of such kink/anti-kink
pairs), in order to be continuous in a compactified
extra dimension. That kind of profile leads to
interesting consequences since in addition to
the ``usual'' brane, where the SM world resides,
there necessarily exists at some distinct point of the extra dimension
another brane, a mirror brane.
In the mirror brane there live particles whose gauge interactions are identical to
those of the SM particles but with reversed chiralities, that is, they are
mirror particles \cite{29}. Besides gravity, the ordinary brane and the mirror
brane can communicate with each other only via possible inert neutrinos, that
is, the right-handed ordinary neutrinos $\nu_{lR}$ and the left-handed mirror
neutrinos $N_{lL}$ that live in the bulk.

We will study in this paper the neutrino sector
of the brane-mirror brane scenario.
We neglect for simplicity any flavour mixing and consider neutrino masses and
mixing within a single family of neutrinos and mirror neutrinos. The neutrino
sector under study thus consists of the fields  $\nu_{L}$, $\nu_{R}$, $N_{L}$
and $N_{R}$. In Sec.\ 2 a concrete realization of our scheme is constructed.
In its framework we derive an effective Lagrangian from which the neutrino
mass matrix is obtained. Unlike in many previous works, our mass matrix is
finite since we integrate out the massive Kaluza-Klein modes.
In Sec.\ 3 we study numerically the generic constraints neutrino data set upon the
various fundamental parameters of the model, such as the size of the extra
dimension, the mass of the sterile neutrino the model predicts and the
fundamental mass scale (the vev of the Higgs doublet) of the mirror brane.
Sec.\ 4 is devoted to discussion of the obtained results. Concluding remarks
are presented in Sec.\ 5.

\section{Theory}

Let us now construct our model for a two-brane world, where one of the branes
traps the SM fermions and the other one the mirror fermions. The gauge inert
states, the right-handed neutrino $\nu_R$ and the left-handed mirror neutrino
$N_L$, are assumed to propagate freely in the whole higher-dimensional bulk.
For the localization mechanism of the gauge-active fermions, based on
effective field theory approach, we follow Refs.\ \cite{26,27}. For simplicity
and clarity we restrict ourselves in the following to the case where there is
only one extra dimension, but the extension of the analysis to the cases of
several extra dimensions is quite straightforward.

We assume the extra dimension  to be compactified on a circle
of radius $R$. The coordinate system
reads $z=(x^\mu ,y)$, where $\mu =0,1,2,3$ and
$y \sim y+2 \pi R$. The scalar field responsible for the
localization of  fermions on the 4-dimensional membranes is denoted by $\Phi$.
It is assumed to have two kinks, one at $y=0$ where its value grows from $-f$
to $+f$ and another one at $y=y_\ast$ where its value drops back to $-f$.
Between the kinks $\Phi$ is supposed to remain constant. Fermions are
localized on a finite-width wall around the points in the fifth dimension
where $\Phi=0$ as, heuristically speaking, their position-dependent masses are
there the smallest. Outside these positions field fluctuations are strongly
suppressed due to  higher masses \cite{27}.

The five-dimensional spinors
are decomposed as
\begin{equation}
  L= \left( \begin{array}{c} N'_R \\ \nu'_L
            \end{array}  \right) \ , \ \ \
  \Psi = \left( \begin{array}{c} \nu'_R \\ N'_L
            \end{array}  \right) \ ,
\end{equation}
where $\nu'_L $ and $\nu'_R$ are related to the ordinary
neutrinos and  $N'_L$ and $N'_R$ to mirror neutrinos. These fields are
functions of the five-dimensional space-time. The appropriateness
of our notation will become clear  towards the end of
this section.

We assume the relevant part of the brane-world action to be of the form
\begin{equation}
   S=\int d^4 xdy[\bar{L} (i\Gamma^A \partial_A
        -m +\Phi )L+i\bar{\Psi} \Gamma^A \partial_A
       \Psi + (\kappa H^\ast \bar{\Psi} L+h.c.)]\ ,
\end{equation}
where $A=0,\ldots ,4$, $m$ is a mass parameter,
$\kappa$ is a dimensionful Yukawa coupling, and $H$
is the neutral component of the standard SU(2)-doublet Higgs field.
The gamma matrices are given by
$$\Gamma^\mu =\left( \begin{array}{cc} 0 & \sigma^\mu
      \\ \bar{\sigma}^\mu & 0 \end{array} \right), \ \
\Gamma^4 =\left( \begin{array}{cc} i & 0
      \\ 0 & -i \end{array} \right),
$$
and all the fields
are taken to be periodically continuous, \emph{e.g.}\
$L(x,0)=L(x,2 \pi R)$. The term $\bar{L}\Phi L $ takes
care of the localization
of the active neutrinos on the branes. In general the action could also
include Dirac and Majorana mass terms for bulk neutrinos \cite{24}, but we
neglect them in the present discussion. It should also be remarked that the
presence of the inert neutrinos in the bulk is an assumption, because we have
omitted in the action the term $\bar{\Psi} \Phi \Psi$, which would have
localized the inert neutrinos on the branes.

The equation of motion for $L$ is
\begin{equation}
    i\Gamma^A \partial_A L=(m -\Phi)L\ ,
\end{equation}
where a term $\kappa^*H\Psi$ has been neglected
as in classical approximation $\Psi=0$.
Supposing that $\Phi =\Phi (y)$, and writing
$N'_R =N_R(x)g_R(y)$ and $\nu'_L =\nu_L(x)g_L(y)$,
one ends up with
\begin{eqnarray}
   ig_L(y) \sigma^\mu \partial_\mu \nu_L & = &
   N_R(\partial_y +m - \Phi (y))g_R(y)\ ,  \nonumber \\
   ig_R(y) \bar{\sigma}^\mu \partial_\mu N_R & = &
   \nu_L(-\partial_y +m - \Phi (y))g_L(y)\ . \label{be1}
\end{eqnarray}
The left-hand side of Eqs.\ (\ref{be1}) being classically zero,
it is trivial to see that $g_{L,R}(y)$ have solutions
behaving so that $g_R(y)$ tends to confine near $y=y_\ast$
and $g_L(y)$ near $y=0$. In other words, since $\nu_L$
and $N_R$ really get localized on different branes,
the chosen action reproduces exactly the features
that we have been looking for.
The consistency of the theory requires
that $\int_0^{2\pi R} dy (\Phi(y)-m)=0$,
which for thin branes reduces to
$\pi -\Delta\theta=\pi m/f$,
where $\Delta\theta= 2\pi -y_*/R$.
Slightly more contrived considerations, which we shall not present
here, reveal that the Higgs field has a twin-peaked
profile with maxima at $y=0$ and $y=y_\ast$. One
may thus approximate (somewhat symbolically,
\emph{cf.}\ the appendix of Ref.\ \cite{6})
\begin{equation}
  \nu'_L =\sqrt{\delta (y)} \nu_L(x)\ , \ \
  N'_R= \sqrt{\delta (y-y_\ast)} N_R(x)\ , \ \
  H=\sqrt{\delta (y)} h_- (x)+ \sqrt{\delta (y-y_\ast)}
   h_+ (x)\ ,
\end{equation}
where $h_- (h_+)$ can be viewed as a vev of the Higgs
field in ``our'' brane (the mirror brane).
By substituting these, together with the familiar
Kaluza-Klein expansion
\begin{equation}
   \left( \begin{array}{c} \nu'_R (x,y) \\
     N'_L (x,y) \end{array} \right)
   = \frac{1}{\sqrt{2 \pi R}} \sum_{n=-\infty}^{\infty}
   \left( \begin{array}{c} \nu_{Rn} (x) \\
     N_{Ln} (x) \end{array} \right) e^{iny/R}\ ,
\end{equation}
to the original action, one has finally
\begin{eqnarray}
  S & = & \int d^4 x \left\{ iN_{R}^{\dagger} \bar{\sigma}^{\mu}
   \partial_\mu N_R +i\nu_{L}^{\dagger} \sigma^{\mu}
   \partial_\mu \nu_L + \sum_{n=-\infty}^{\infty}
   \left[ i\nu_{Rn}^{\dagger} \bar{\sigma}^{\mu}
   \partial_\mu \nu_{Rn} +iN_{Ln}^{\dagger} \sigma^{\mu}
   \partial_\mu N_{Ln}  \right. \right. \nonumber \\
    & & \left. \left. +\frac{in}{R}(\nu_{Rn}^{\dagger} N_{Ln}
     -N_{Ln}^{\dagger} \nu_{Rn})+u(h_{-}^{\ast}
     \nu_{Rn}^{\dagger} \nu_L +h_{+}^{\ast}
     N_{Ln}^{\dagger} N_R \, e^{in \Delta \theta}
     +h.c.)\right] \right\} \ ,        \label{b1}
\end{eqnarray}
where
\begin{equation}
u=\frac{\kappa}{\sqrt{2 \pi R}} \ , \label{uu}
\end{equation}
and $\kappa$ is taken to be real.

The first six $n$-dependent terms of the action (\ref{b1})
disappear due to the equations of motion of the Kaluza-Klein
excitations $\nu_{Rn}$ and $N_{Ln}$ (with $n \neq 0$),
\begin{eqnarray}
   N_{Ln} & = & \frac{iR}{n} (i \bar{\sigma}^{\mu}
   \partial_\mu \nu_{Rn} +uh_{-}^{\ast} \nu_L)\ , \nonumber \\
   \nu_{Rn} & = & -\frac{iR}{n} (i \sigma^{\mu}
   \partial_\mu N_{Ln} +uh_{+}^{\ast} N_R \, e^{in \Delta \theta})\ .
   \label{b2}
\end{eqnarray}
The remaining two terms of the action then yield
\begin{eqnarray}
  \mathcal{L}_{n \neq 0} & = & \sum_{n \neq 0}
  \left[ \left(iu^2 \frac{R}{n} h_+ h_{-}^{\ast} N_{R}^{\dagger}
   \nu_L \, e^{-in \Delta \theta} +h.c. \right) +u\frac{R}{n}
   h_- \nu_{L}^{\dagger} \sigma^{\mu} \partial_{\mu}
   N_{Ln} \right. \nonumber  \\  & & \left.
   -u\frac{R}{n}h_+ N_{R}^{\dagger} \bar{\sigma}^{\mu}
   \partial_{\mu} \nu_{Rn} \, e^{-in \Delta \theta} \right] \ ,
\end{eqnarray}
or, by applying Eqs.\ (\ref{b2}) again,
\begin{eqnarray}
  \mathcal{L}_{n \neq 0} & = & u^2 s(\Delta \theta) R
   (h_+ h_{-}^{\ast} N_{R}^{\dagger} \nu_L +h.c.)
   + \sum_{n \neq 0} iu^2 \frac{R^2}{n^2}
   h_- \nu_{L}^{\dagger} \sigma^{\mu} \partial_{\mu}
   (h_{-}^{\ast} \nu_L) \nonumber \\ & &
   + \sum_{n \neq 0} iu^2 \frac{R^2}{n^2}
   h_+ N_{R}^{\dagger} \bar{\sigma}^{\mu} \partial_{\mu}
   (h_{+}^{\ast} N_R)\ ,   \label{b3}
\end{eqnarray}
where terms including higher derivatives of
$\nu_{Rn}$ and $N_{Ln}$ have been neglected, and
\begin{equation}
  s(\Delta \theta)=i\sum_{n \neq 0}
  \frac{e^{-in \Delta \theta}}{n}
  =\pi -\Delta \theta \ \ (\Delta \theta \neq 0)\ .
\end{equation}
This on-shell Lagrangian could be equivalently
obtained by integrating out the massive  Kaluza-Klein
excitations.

Note that since $h_{\pm}^{\ast}$ are constants
in the leading order, they can well be taken out
of the derivatives in Eq.\ (\ref{b3}).
Examinations show that the
values of $h_\pm$ are essentially dependent on the widths of the
ordinary and mirror branes. The quantity $h_+$ can be considered here as
a free parameter while $h_-$ is bound by the usual Higgs scalar
expectation value $|h_-|=174$ GeV.

Combining finally Eq.\ (\ref{b3})
and Eq.\ (\ref{b1}) (for $n=0$) with suitable
rescalings of the fields so that the kinetic terms take the
canonical form, one ends up with a  Lagrangian
\begin{eqnarray}
  \mathcal{L}_{eff} & = & iN_{R}^{\dagger} \bar{\sigma}^{\mu}
   \partial_\mu N_R +i\nu_{L}^{\dagger} \sigma^{\mu}
   \partial_\mu \nu_L
   + i\nu_{R0}^{\dagger} \bar{\sigma}^{\mu}
   \partial_\mu \nu_{R0} +iN_{L0}^{\dagger} \sigma^{\mu}
   \partial_\mu N_{L0} \\ & &
    +u^2 s(\Delta \theta) R
   (n_+ n_- h_+ h_{-}^{\ast} N_{R}^{\dagger} \nu_L +h.c.)
   +u(n_- h_{-}^{\ast} \nu_{R0}^{\dagger} \nu_L
   +n_+ h_{+}^{\ast} N_{L0}^{\dagger} N_R +h.c.)
    \nonumber \ ,
\end{eqnarray}
where
$$n_{\pm}=\frac{1}{\sqrt{1+\frac{\pi^2}{3}u^2 R^2 |h_{\pm}|^2}}\ ,
$$
and the identity $\sum_{n \neq 0} n^{-2}=\pi^2 /3$
has been used. In this Lagrangian, where all the massive
Kaluza-Klein modes are integrated out, only the lowest order
effects of the extra dimension
(\emph{i.e.}\ field rescalings and mass terms) are kept.

\section{Neutrino phenomenology}

According to the Lagrangian $\mathcal{L}_{eff}$ the neutrino mass term
of the model is given by
\begin{equation}
  \mathcal{L}_M =-(\overline{\nu}_L\, \overline{\nu}_R\,
    \overline{N}_R\, \overline{N}_L) \left(
  \begin{array}{cccc}
    0 & a & b & 0  \\
    a & 0 & 0 & 0  \\
    b & 0 & 0 & c  \\
    0 & 0 & c & 0  \end{array} \right)
  \left(  \begin{array}{c}
     \nu_L \\ \nu_R \\ N_R \\ N_L
     \end{array} \right)\ ,   \label{b4}
\label{massmatrix}\end{equation}
where
\begin{equation}
  a=un_- |h_{-}|\ , \ \ b=u^2 s(\Delta \theta) R
   n_+ n_- |h_+ h_{-}|\ , \ \   \label{abc}
  c=un_+ |h_{+}|\ ,
\end{equation}
with appropriate redefinitions of the fields,
and we have associated $\nu_R$ and $N_L$ with the zero modes
of the respective Kaluza-Klein states. More familiarly the mass
Lagrangian can be presented in the form
\begin{equation}
    \mathcal{L}_M =-\frac{1}{2} \overline{\mathcal{N}}_{R}^{c}
     M \mathcal{N}_L +h.c.\ ,
\end{equation}
where $\mathcal{N}_{L}^{T} =(\nu_L \, \nu_{L}^{c} \,
N_{L}^{c} \, N_L)$, and $M$ is the matrix appearing in Eq.\ (\ref{b4}).

As one can see from the results above, the dependence of the neutrino
mass Lagrangian on the fundamental parameters of the theory, such as the
radius $R$, the values of the scalar, $h_{\pm}$, the coupling $\kappa$, and
the relative positions of the brane and mirror brane (\emph{i.e.}
$\Delta\theta$), is quite non-trivial and difficult to analyse analytically.
We will therefore study numerically the generic constraints the present
neutrino data sets on the model  by varying unknown parameters within
conceivable range of values and plotting the ensuing predictions for various
measurable quantities.

Let us start the phenomenological analysis by diagonalizing the neutrino
mass matrix obtained above. Since $M$ is real and symmetric, the
diagonalization can be performed by an orthogonal matrix $O$.
(We consider here just one neutrino family. The different families
may have different mixing angles.)

If we define
\begin{equation}
O=\frac{1}{\sqrt{2}} \left( \begin{array}{cccc}
       \sin\alpha & \sin\alpha  & \cos\alpha  & -\cos\alpha  \\
       -\sin\beta & \sin\beta   & -\cos\beta  & -\cos\beta   \\
       -\cos\beta & \cos\beta   & \sin\beta   & \sin\beta    \\
       \cos\alpha & \cos\alpha  & -\sin\alpha & \sin\alpha
     \end{array}  \right)\ ,   \label{c1}
\end{equation}
with
\begin{equation}
\sin^2 \alpha = \frac{1}{2} +
 \frac{a^2 +b^2 -c^2}{2\sqrt{(a^2 +b^2 +c^2)^2 -4a^2 c^2}}\ , \ \
\sin^2 \beta  = \frac{1}{2} +
 \frac{a^2 -b^2 -c^2}{2\sqrt{(a^2 +b^2 +c^2)^2 -4a^2 c^2}}\ ,
            \label{c2}
\end{equation}
we obtain
\begin{equation}
O^T MO=\left( \begin{array}{cccc}
        -m_+ & 0 & 0 & 0  \\
        0 & m_+  & 0 & 0  \\
        0 & 0 & -m_- & 0  \\
        0 & 0 & 0 & m_-
     \end{array}  \right)
  \equiv
 \left( \begin{array}{cccc}
        \sigma_1 m_+ & 0 & 0 & 0  \\
        0 & \sigma_2 m_+  & 0 & 0  \\
        0 & 0 & \sigma_3 m_- & 0  \\
        0 & 0 & 0 & \sigma_4 m_-
     \end{array}  \right) \ ,    \label{c3}
\end{equation}
where the eigenvalues are
\begin{equation}   \label{c4}
m_{\pm} = \sqrt{\frac{1}{2} \left(a^2 +b^2 +c^2 \pm
          \sqrt{(a^2 +b^2 +c^2)^2 -4a^2 c^2} \right)}\ ,
\end{equation}
and $\sigma_i$'s are sign factors. Note that if $b=0$,
matrix $O$ can be presented in a
much simpler form. This case, however, is realized only in the
special case of $\Delta \theta =\pi$, \emph{i.e.} when the two branes
are in  opposite locations in the fifth dimension.

The mass eigenstates are given by
\begin{equation}
  \chi_i =\sum_j \left(O^{T}_{ij} (\mathcal{N}_L)_j +
  \sigma_i O^{T}_{ij} (\mathcal{N}^{c}_{R})_j \right)\ ,
         \label{c5}
\end{equation}
in terms of which the mass Lagrangian  reads
\begin{equation}
  \mathcal{L}_M =-\frac{1}{2}m_+
   (\bar{\chi}_1 \chi_1 + \bar{\chi}_2 \chi_2)
        -\frac{1}{2}m_-
   (\bar{\chi}_3 \chi_3 + \bar{\chi}_4 \chi_4)\ .
\end{equation}
By defining
\begin{equation}    \label{c7}
  \psi =\frac{1}{\sqrt{2}}(\chi_1 +\chi_2)\ , \ \
  \phi =\frac{1}{\sqrt{2}}(\chi_3 -\chi_4)\ ,
\end{equation}
one has finally
\begin{equation}
   \mathcal{L}_M =-m_+ \bar{\psi} \psi
                  -m_- \bar{\phi} \phi \ .
\end{equation}
The neutrino sector thus consists of two Dirac neutrinos, as was
expected as the theory conserves, by construction, lepton number.

  From Eqs.\ (\ref{c1}),
(\ref{c5}) and (\ref{c7}) it follows that the ordinary left-handed neutrino
$\nu_L$ is the following superposition of the mass eigenstate neutrinos:
 $$
\nu_L=\cos \alpha\ \phi_L +\sin \alpha\ \psi_L.
$$
The superposition orthogonal to this combination is the inert field $N_L$.
As $\nu_L$ is the only active neutrino living in our brane, only the mixing
angle $\alpha$ can be experimentally probed.  The other angle $\beta$
parametrizes the mixing between the right-handed fields $\nu_R$ and $N_R$, and
 is therefore a measurable quantity only in the mirror world.

There exist several empirical constraints on the masses and mixing angles of
neutrinos, coming from laboratory experiments, astrophysical observations and
cosmological considerations. In our case, where flavour mixings are neglected,
mixings occur between an active and a sterile neutrino of each family. It
turns out that the existing laboratory bounds on the mixing angle $\alpha$
between the electron neutrino and a heavier neutrino in various mass ranges of
the heavier neutrino mass are in general ineffective in our model. This is
because the mass of the extra neutrino turns out to be quite low for a
plausible choice of the model parameters. The only relevant
laboratory constraints are due to the electron neutrino
mass measurement in the tritium beta decay
experiments \cite{70} and the neutrino oscillation
experiments. The upper limit of the electron neutrino
mass from the beta decay experiment is \cite{70}
\begin{equation}
m_{\nu_{e}} \lsim 2.3\ {\rm eV} .
\end{equation}
As far as the electron neutrino is concerned the most stringent
oscillation limit comes from the Bugey disappearance experiment
\cite{70b}.
This limit has been approximated as follows in our numerical
calculation:
\begin{equation}
\begin{array}{ll}
    \sin^2 2\alpha < 0.1
    & \mbox{for}\  100\ {\rm eV}^2> \vert \delta m^2\vert > 2\ {\rm eV}^2\ ,
      \\
    \sin^2 2\alpha < 0.02
    & \mbox{for}\  2\ {\rm eV}^2> \vert \delta m^2\vert > 0.04\ {\rm eV}^2\ ,
      \\
    \sin^2 2\alpha < 0.1
    & \mbox{for}\  0.04\ {\rm eV}^2> \vert \delta m^2\vert > 0.01\ {\rm eV}^2
    \ .  \label{bugey}
\end{array}
\end{equation}

The active-sterile mixing can also be constrained by cosmological arguments.
If the mixing is too large, neutrino oscillations, acting as an
effective interaction, would bring the sterile neutrino in equilibrium
before neutrino decoupling, and the resulting excess in energy density
would endanger the standard scheme for the nucleosynthesis of light
elements (BBN) \cite{71}. This leads to the following bound for
$\nu_e \leftrightarrow \nu_s$ mixing \cite{72}:
\begin{equation}
\begin{array}{ll}  \label{shi}
    |\delta m^2| \sin^2 2\alpha <5 \times 10^{-8}\
    \mbox{eV}^2 & \mbox{for}\ |\delta m^2|<4\ \mbox{eV}^2\ ,
      \\
    \sin^2 2\alpha <10^{-8}\
    & \mbox{for}\ |\delta m^2|>4\ \mbox{eV}^2\ .
\end{array}
\end{equation}
This bound is, however, avoided if there was a suitable net lepton number
in the early universe \cite{73}.

With applying these constraints, we
search the allowed regions of the parameter space numerically using a
Monte Carlo analysis where
we have varied three unknown parameters:
the radius of the extra dimension $R$, the mirror brane Higgs
value $h_+$ and the mirror brane position parameter
$\Delta\theta$. The extra dimension radius is varied from the
Planck scale to a millimeter scale and the
mirror brane Higgs value between $10^{-5}$ MeV and $10^{10}$ MeV.
We consider these parameter ranges wide enough for a representative analysis.
Both $R$ and $h_+$ are randomized so that their logarithms are
evenly distributed. For the position parameter $\Delta\theta$
an even distribution between 0 and $\pi$ has been taken.

In addition to these parameters the action depends on the dimensionful
coupling $\kappa$. The natural scale for it could possibly be set
by the higher-dimensional Planck scale $M_*$, and therefore we
write (with $M_{Pl}^2 =M_{\ast}^{3} R$)
\begin{equation} \label{kappa}
\kappa=\frac{\kappa '}{\sqrt{M_*}}=
 \kappa^\prime \left ({R\over M_{Pl}^2}\right )^{1/6},
\end{equation}
where $\kappa^\prime$ is a dimensionless coupling constant.
Its value is not really known, but a plausible choice would
be a number relatively close to unity. We have performed our analysis for  the
values $\kappa^\prime = 1$ and $\kappa^\prime = 0.01$.

A central parameter in the scheme is obviously the size $R$ of the extra
dimension. From the experiments that probe the effects of gravity in short
distances one knows that \cite{4} \begin{equation}
R\lsim 1\ {\rm mm}.
\label{Rlimit}
\end{equation}
  From the dependence of the elements of the mass matrix $M$ on $R$ follows
an upper bound on the larger mass eigenvalue $m_+$  as a function of $R$, as
presented in Fig.\ 1. When the limit (\ref{Rlimit}) is saturated, the largest
allowed value is $m_+ \sim 10$ eV, but larger values are allowed for a smaller
$R$.

\begin{figure}[ht]
\leavevmode
\centering
\vspace*{60mm}
\begin{picture}(0,60)(0,490)
\includegraphics{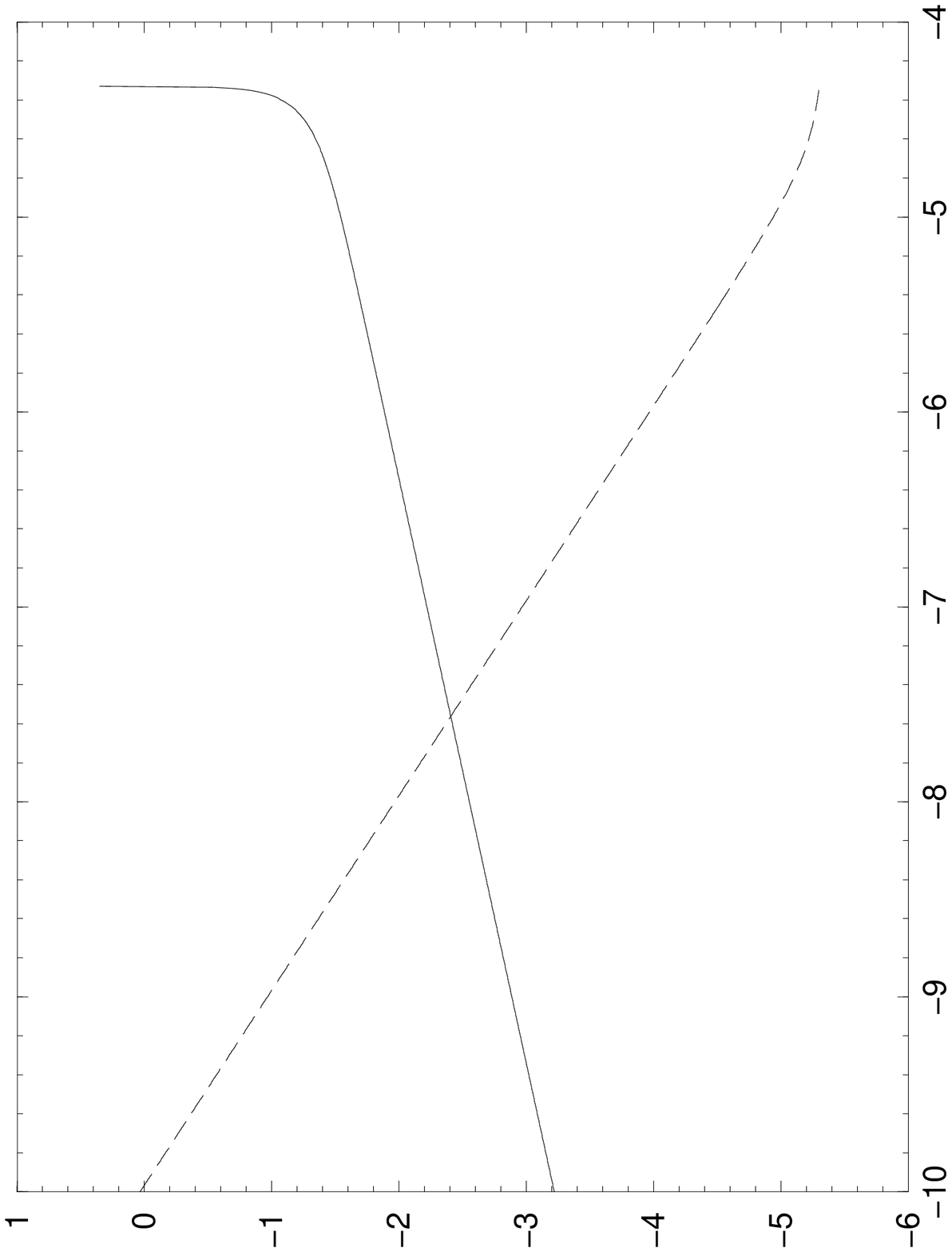}
\put(-200,460){$\scriptstyle \log({m_+^{\rm max}\over {\rm MeV}})$}
\put(-190,435){$\scriptstyle \log (\kappa ')$}
\put(-18,325){$\scriptstyle \log({R_{\rm min}\over {\rm mm}})$}
\end{picture}
\figcap{Maximal neutrino mass $m_+^{\rm max}$(dashed line) and coupling
$\kappa '$ (solid line) as function of minimal radius of the extra dimension
$R_{\rm min}$.
  \label{kuva1}
}
\end{figure}


\section{Results and discussion}

Let us now proceed to our numerical results.
In Fig.\ 2 we present a scatter plot in the $(R,\sin\alpha)$-plane,
obtained by
allowing the model parameters vary as indicated above, with and without
taking the cosmological constraint, Eq.\ (\ref{shi}), into account (marked in
the plot by crosses and dots, respectively). Fig.\ 2a corresponds to the case
$\kappa'=0.01$. Let us now look at the general features displayed by this
figure. Comparison of Eqs.\ (\ref{uu}), (\ref{abc}) and (\ref{kappa}) implies
that $a,c \gg b$ for small values of $R$. From Eq.\ (\ref{c2}) it can then be
inferred that, depending on the relative sizes of $h_+$ and $h_-$,   either
$\sin ^2\alpha\simeq 1$ ($h_+<h_-$) or  $\sin ^2\alpha\simeq 0$ ($h_+>h_-$).
In other words, for small values of the size $R$ of the extra dimension, the
mixing angle $\alpha$ has values either close to 0 or close to $\pi/2$,  the
other values being forbidden. The small values of  $\alpha$ correspond to the
case where $a<c$, that is, where the ordinary active neutrino $\nu_L$ is
predominantly the lighter mass eigenstate $\phi_L$ and the sterile mirror
neutrino $N_L$ is predominantly the heavier state $\psi_L$. The values of
$\alpha$ close to $\pi/2$ in turn correspond to the situation where $a>c$, and
  $\nu_L$ is predominantly the heavier state $\psi_L$ and the sterile mirror
neutrino $N_L$ is predominantly the lighter state $\phi_L$, the both neutrino
masses now being below the experimental limit $m_{\pm}\lsim 2.3$ eV.

\begin{figure}[ht]
\leavevmode
\centering
\vspace*{100mm}
\begin{picture}(0,60)(0,490)
\includegraphics{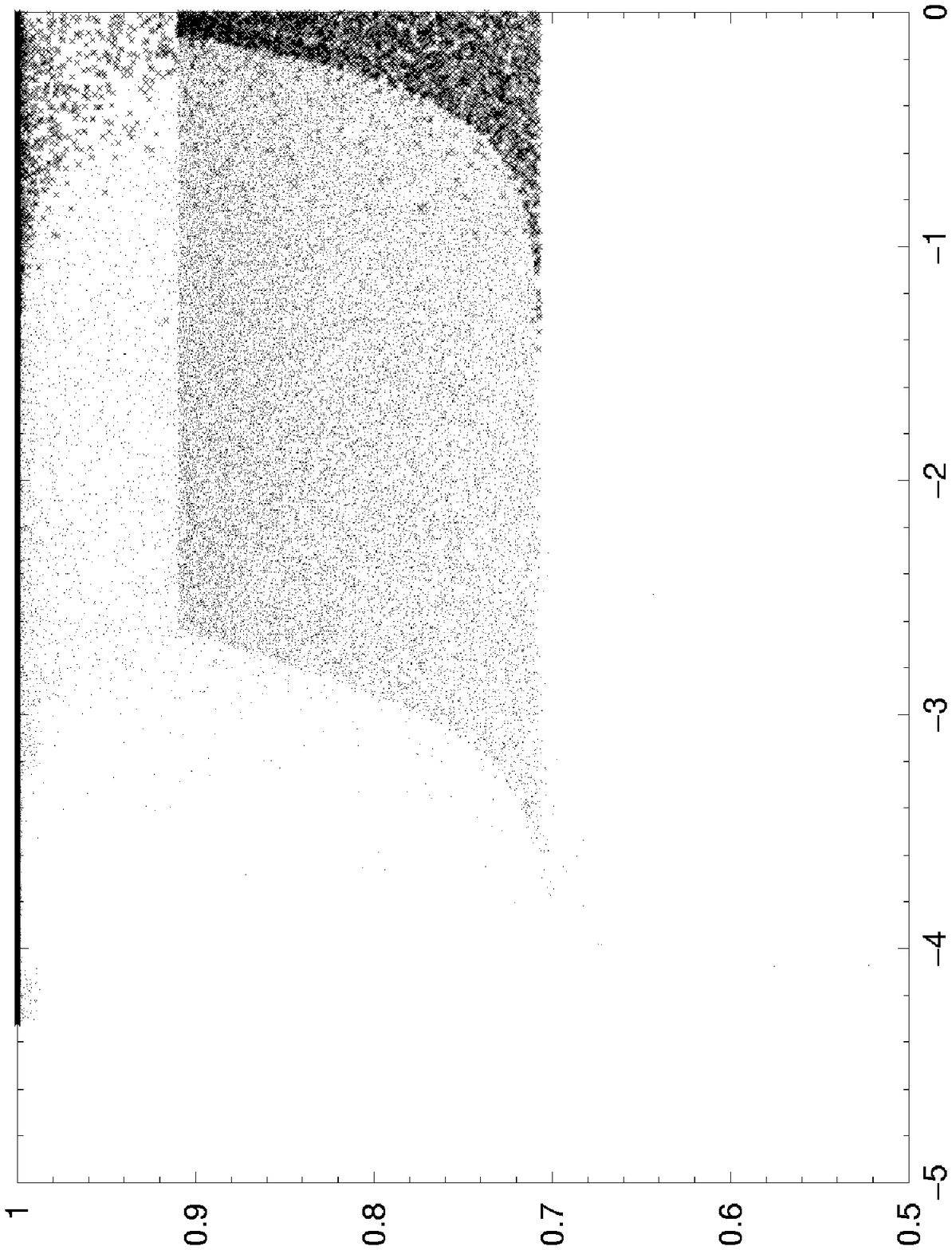}
\put(-180,320){$\scriptstyle \sin \alpha$}
\put(-10,225){$\scriptstyle \log ({R\over{\rm mm}})$}
  \put(-120,385){(b)}
\includegraphics{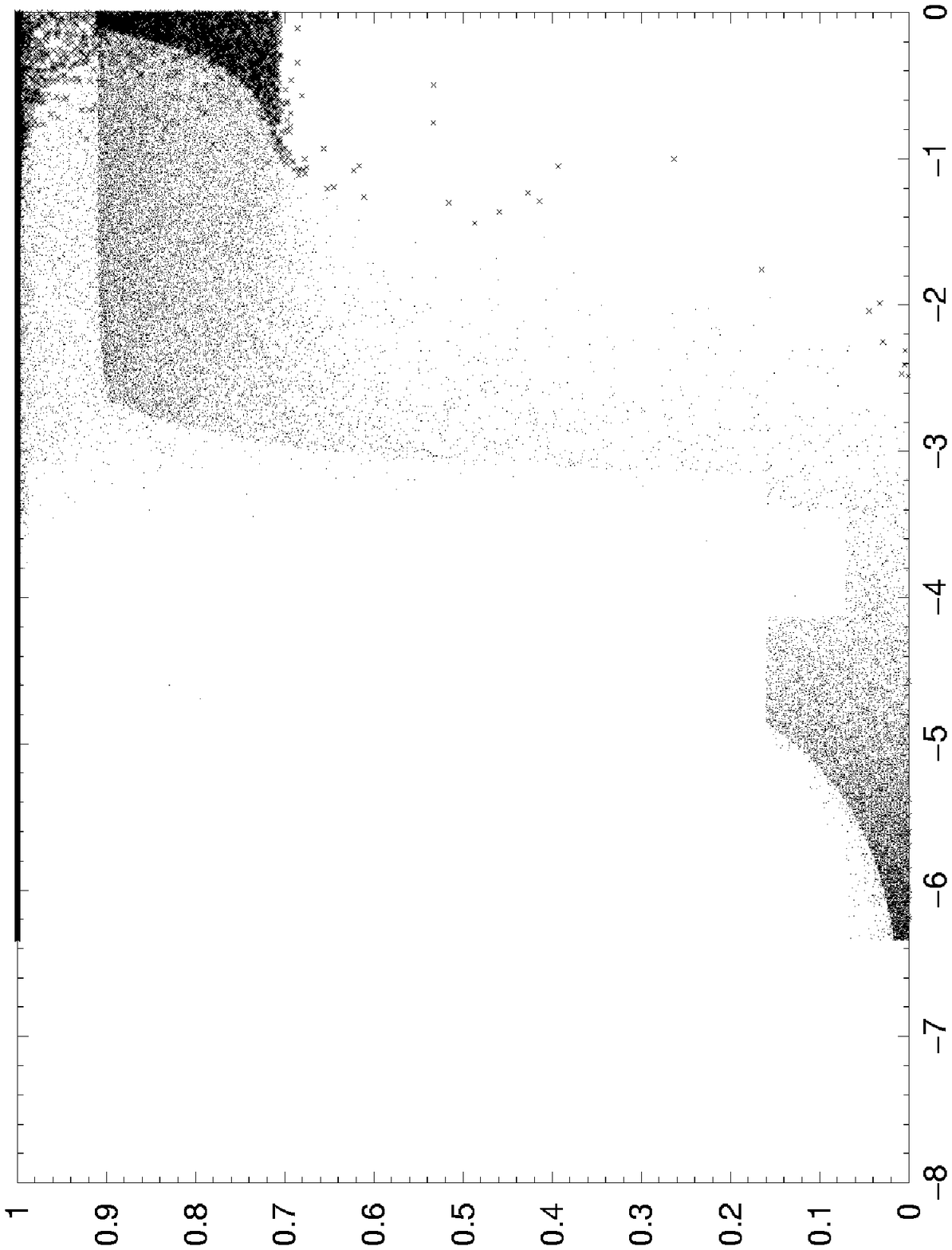}
\put(-180,520){$\scriptstyle \sin \alpha$}
\put(-10,425){$\scriptstyle \log ({R\over{\rm mm}})  $}
 \put(-120,585){(a)}
\end{picture}
\figcap{The neutrino mixing angle $\sin (\alpha )$ as function of extra
dimension radius $R$ for $\kappa'=0.01$ (a) and $\kappa'=1.0$ (b). Crosses
indicete points allowed both oscillations and cosmology, whereas dot are
points allowed by oscillations only.
  \label{kuva2}
}
\end{figure}


It is important to note that the size of the extra dimension, $R$,
is bounded from
below in our scenario. This is an obvious consequence of the fact that the
quantities $a$ and $c$ increase with decreasing $R$, and the mass of the
predominantly active neutrino state $\simeq \nu_L$, whose mass experiments
test, is determined mainly by these quantities, since for small $R$ $a,c\gg
b$.  In the case of $\kappa'=0.01 (1.0)$ the lower limit is $R\gsim 4 \times
10^{-7} (5 \times 10^{-5})$ mm. More generally
the lower limit of $R$ is obtained from
\begin{equation}
\frac{\kappa ' h_-}{\sqrt{2 \pi}}<(R M_{Pl})^{1/3} \times
2.3\ \mbox{eV}\ ,
\end{equation}
as long as $\kappa ' \lsim 0.1$ (\emph{cf.}\ the solid line
in Fig.\ 1).

When the cosmological constraint, Eq.\ (\ref{shi}), is taken into account
(crosses in Figs.\ 2a and 2b), large values of $R$ are favoured
($R\gsim 3 \times 10^{-2}$ mm),
unless the mixing angle $\alpha$ is very close to 0 or $\pi/2$. This is
because for smaller $R$ the squared mass difference $ m_+^2-m_-^2$ tends to
increase, jeopardizing the  fulfillment of the cosmological condition. Let us
note that in the parameter space covered by our Monte Carlo analysis, the
cases allowed by the cosmological condition are concentrated to the region of
large mixing angle $\alpha$, \emph{i.e.}\ to the case where  the active
neutrino is heavier than the sterile one. Nevertheless, also the small values
of the mixing angle are allowed, \emph{i.e.}\ the case where the sterile
neutrino can be relatively heavy, albeit with quite a small part of the
parameter space.

The step-like behavior of the scatter plot in the region
$10^{-5}$ mm $\lsim R\lsim 10^{-3}$ mm
is due to the oscillation constraint, Eq.\ (\ref{bugey}).
As remarked before, this bound together with the
upper bound of 2.3 eV
of the electron neutrino mass, is the only
laboratory constraint that is effective for our model.
The reason is the relatively small values of
the inert neutrino mass the theory allows,
as  displayed by Fig.\ 1.

Fig.\ 2b is the same scatter plot as in Fig.\ 2a,
except that now $\kappa'=1$.
A comparison with the previous case reveals the interesting fact, namely that
for a large values of the coupling $\kappa'$ only the large mixing angles
$\alpha$ are allowed. This can be understood as follows. The ratio of the mass
matrix elements $a$  and $c$,
\begin{equation} \frac{a}{c}= \frac{n_-\vert h_-
\vert}{n_+\vert h_+ \vert}=
\frac{|h_-|}{|h_+|}
\sqrt{\frac{1+\pi^2u^2R^2 |h_+|^2/3}{1+\pi^2u^2R^2 |h_-|^2/3}}\ ,
\end{equation}
is approximately equal to 1 when the product $u^2R^2$ is large
enough. From Eq.\ (\ref{c2}) one can deduce that in this case $\sin^2\alpha
\gsim 1/2$. In the case $\kappa'=1$ one is practically always in this regime,
taken into account the lower limit for $R$ from the electron neutrino mass
measurement. In the previous case of $\kappa'=0.01$ this regime is reached,
because of the smaller value of $u$, only at the upper end of the allowed $R$
range, as can been seen in Fig.\ 2a. The lower bound for $R$, still given by
the experimental upper limit for the electron mass, is now larger than in the
case of $\kappa'=0.01$ since the neutrino mass (for small $b$) is proportional
to $\kappa'$, as seen from Eqs.\ (\ref{abc}), (\ref{c4}) and (\ref{kappa}).
When $R$ is increased, all possibilities from $\nu_L\simeq \psi_L$ to the
maximal mixing $\nu_L=(\phi_L+\psi_L)/\sqrt{2}$ become open. At low $R$
cosmology forces $\nu_L$ to be very close to the heavier mass state $\psi_L$,
but when $R\gsim 3 \times 10^{-2}$ mm, $\nu_L$ may develop a large $\phi$
component as well.

\begin{figure}[ht]
\leavevmode
\centering
\vspace*{65mm}
\begin{picture}(0,60)(0,490)
\includegraphics{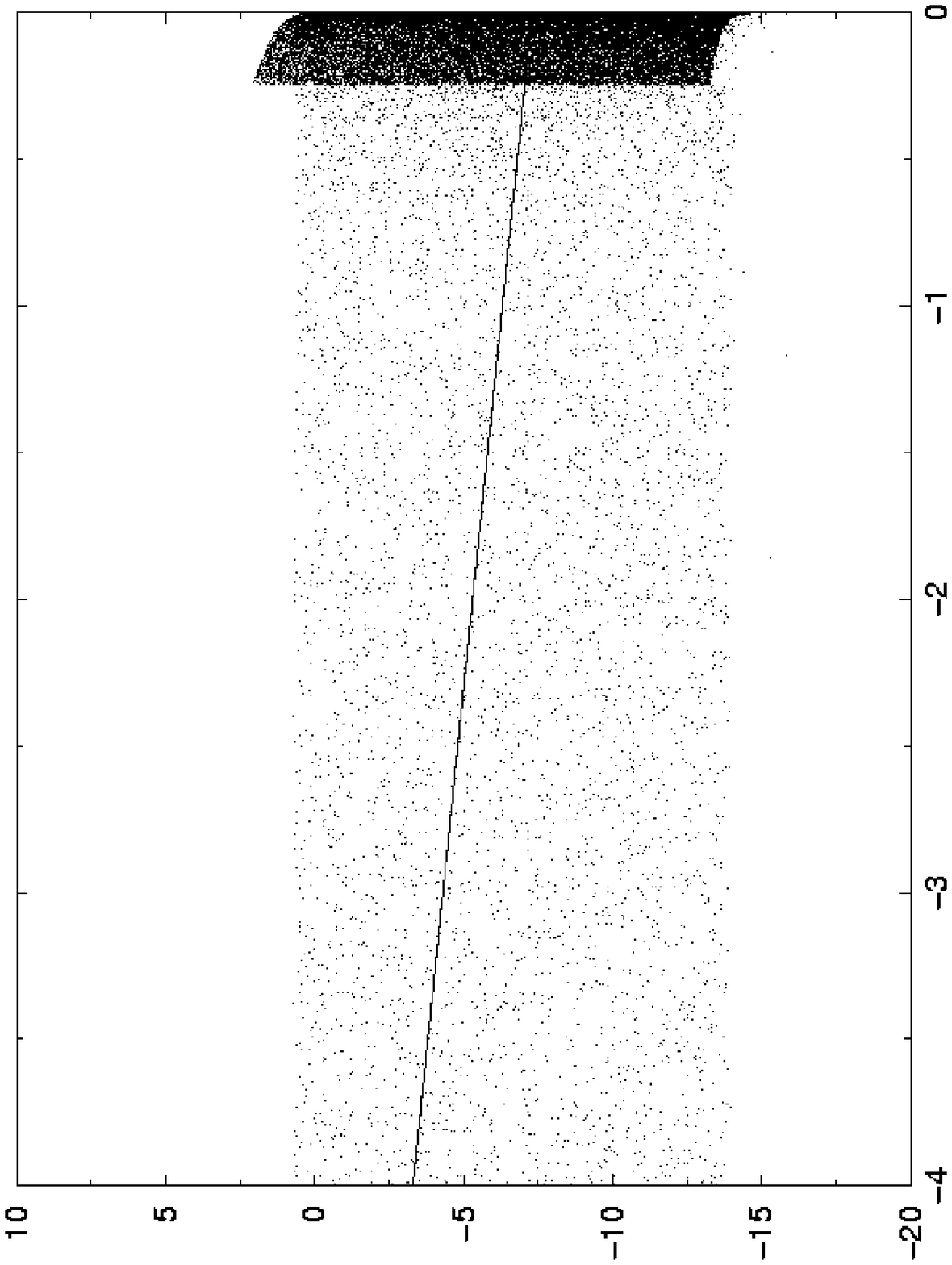}
\put(-200,440){$\scriptstyle \log ({\delta m^2\over {\rm MeV}^2})$}
\put(-18,310){$\scriptstyle \log (\sin^2(2\alpha))$}
\end{picture}
\figcap{Neutrino mass difference as a function of $\sin^2(2\alpha )$ . The
solid line divides the space to excluded (upper part) and allowed (lower part)
regions by cosmological considerations ($\kappa'=1.0$).
  \label{kuva3}
}
\end{figure}


Fig.\ 3 presents the region in the parameter space
$(\sin^2 2\alpha , \delta m^2)$
(with $\delta m^2 =  m_+^2 - m_-^2$)
reached by our model in the case $\kappa'=1$.
The only constraint taken into
account here is the upper bound 2.3 eV for electron neutrino mass.
We have checked that the inclusion of other laboratory bounds do
alter Fig.\ 3, but only cuts $\sim 1$\% more points from
the region which is anyway excluded by oscillation experiments.
The line between the cosmologically allowed (lower part)
and excluded (upper part) regions has been drawn to Fig.\ 3, too.
For $\kappa'=0.01$ the
allowed region would extend to larger mass differences, with virtually no
change in cosmologically allowed region. As one can see, our model allows for
all the active-sterile mixing schemes discussed as possible solutions to the
solar neutrino and atmospheric neutrino anomalies.
In particular the so-called SMA solution ($|\delta m^2| \approx (3-10) \times
10^{-6}\     \mbox{eV}^2, \ \sin^2 2\alpha \approx (0.2-1.3)     \times
10^{-2}$) of the solar neutrino anomaly can be realized in this model. Note,
however, that the solar LMA solution is excluded by cosmology bound.

\section{Conclusions}

We have investigated a five-dimensional model
with two four-dimensional branes, a brane
for ordinary SM particles and another brane for mirror particles.
By looking at neutrino masses and mixings, we have shown that confrontation
of the predictions of the model with data constrains the extra
dimension physics remarkably. The experimental upper bounds
on the masses of the known neutrinos directly
restrict the extra dimension size,
independently of how neutrinos actually mix.
The value of brane-bulk neutrino
coupling $\kappa '$ determines the minimal value of extra
dimension size $R$, a larger
coupling corresponding to a larger $R_{\rm min}$.
The nature of neutrino mixing depends crucially on
the coupling $\kappa '$. For $\kappa '\sim 1$
the predominantly active neutrino, \emph{i.e.}\ the ordinary
neutrino, is never lighter than the predominantly
sterile neutrino. For smaller values, $\kappa '\sim 0.01$,
also the opposite situation is allowed. In both cases active
and sterile neutrino can mix maximally when $R$
is large enough.

As may be seen from Fig.\ 1, also
the largest possible mass $m_+^{\rm max}$
of the heavier neutrino (of each flavour)
depends essentially on $\kappa '$, a smaller
$\kappa '$ (\emph{i.e.}\ a smaller $R_{\rm min}$)
enabling $m_+^{\rm max}$ to be higher. For
$\kappa ' = 0.01$ the maximal neutrino mass is $\sim 1$ keV.
However, already for $\kappa ' \sim 10^{-3}$ the neutrino
mass may be of the order of 1 MeV with $R_{\rm min} \sim 10^{-10}$ mm.
Therefore the issue of the value of $\kappa '$ is emphasized.

If  also cosmological constraints are taken into account, the room for allowed
neutrino mixings is largely suppressed. With cosmological bound the model
clearly favor large radius $R \gsim 10^{-2}$ mm. Only mixing angles
very close to $\pi /2$ (and equally to 0 for some values of
$\kappa '$) make exception, all the radii down to the
minimal one being then allowed. In this case no direct
method exists for observing mirror brane neutrinos.

The model we have studied is the simplest possible
in the sense that only one extra dimension is introduced.
This kind of approach is not fully realistic
since it is known that the hierarchy
problem cannot be accounted for unless the number of extra
dimensions is at least two. 
It is clear that with more than one extra dimension
the brane-bulk structure
becomes much more complicated than in the five-dimensional case
we have studied in this paper,
increasing the flexibility in the values of the model parameters.
Nevertheless, we expect
that our model exhibits the generic features that also appear
in higher-dimensional treatments.

\section*{Acknowledgments}
This work has been supported by the Academy of Finland
under the project no.\ 40677.

\end{document}